\documentstyle[epsf]{article}
\begin{document}
\pagestyle{plain}

\title{
{\bf Randomness Evaluation and Hardware Implementation of Nonadditive CA-Based Stream Cipher}}

\author{%
Song-Ju \textsc{Kim}$^{1}$ and Ken \textsc{Umeno}$^{1,2}$\footnote{E-mail Addr
ess: \{songju, umeno\}@nict.go.jp}\\
\\
\noindent
$^{1}$\hspace{2mm} National Institute of Information and Communications
Technology, Japan\\
$^{2}$\hspace{2mm} ChaosWare Inc., Japan\\
\\
\Large{\bf Abstract}\\
We shall review the cellular automaton (CA)-based pseudorandom-number\\
 generators (PRNGs), and show that one of these PRNGs can generate 
high-quality\\ random numbers which can pass all of the statistical tests
 provided by the National\\ Institute of Standards and Technology (NIST).
A CA is suitable for hardware imple\\mentation.
We demonstrate that the CA-based stream cipher, which is implemented\\ 
 in the field-programmable gate arrays (FPGA), has a high encryption speed
 in a real\\
\hspace{-70mm}-time video encryption and decryption system.
\hspace{-18mm}FPGA Implementation, Real-time Encryption}

\maketitle

\section{Introduction}

Many secret key cryptosystems have been proposed thus far~\cite{ref:handbook}.
One of the advantages of a stream cipher over a block cipher is 
its high encryption speed and small gate size when it is implemented in hardware.
Therefore, a stream cipher is suitable for developing high-speed and
low-power encryption systems.
There will be an increase in demand for the development of faster
encryption associated with high-resolution video, high-volume data
retrieval, and other high-speed data communication systems such as
10 Gbit networks. 
Thus, in various fields, a hardware-like stream cipher is now 
requested for real-time encryption and decryption. 

In this paper, we propose a CA-based PRNG used for a stream cipher which is suitable for
hardware implementation because of its simple construction (i.e., locality of
interaction and homogeneous units)~\cite{ref:CHES}.
A one-dimensional elementary cellular automaton (ECA) consists of a line
of cells with $S_i$ $=$ 0 or 1 for $i=0,1,2,\cdots,N$.
These cell values are updated in parallel in discrete time steps according to
a fixed rule of the form,
\begin{equation}
S_i^{t+1} = F (S_{i-1}^t, S_i^t, S_{i+1}^t)
\label{eq1}
\end{equation}
where $S_i^t$ denotes the $i$ cell value at time $t$.
Wolfram firstly used the ECA as a PRNG, and investigated its randomness~\cite{ref:wolfram1}. 
He concluded that the following `rule 30' is the best PRNG 
among the ECA rules,
\begin{equation}
S_i^{t+1} = S_{i-1}^t \oplus S_i^t \oplus S_{i+1}^t \oplus S_i^t \cdot S_{i+1}^t
\label{eq2}
\end{equation}
or 
\begin{equation}
S_i^{t+1} = S_{i-1}^t \hspace{1mm}XOR\hspace{1mm} ( S_{i}^t \hspace{1mm}OR\hspace{1mm} S_{i+1}^t )
\label{eq3}
\end{equation}
where $\oplus$ denotes plus modulo 2.
He also proposed the CA-based stream cipher using this ECA30~\cite{ref:wolfram2}.
It is known that ECA30 has large periodic cycles.
The maximum period is $2^{0.6N}$ with systems size $N$.
Most configurations fall into the cycle if we set the system
size $N$ sufficiently large.
Now, Wolfram also emphasizes that ECA30 is the `origin of randomness' in his new
book~\cite{ref:new}.

On the other hand, additive CA-based PRNGs, such as rule 90, rule 150, rule 105, and
rule 165,\footnote{Rules of linear CA have only XORs, and rules of additive
CA have only XOR and XNOR.} have been proposed by 
Hortensius et al.~\cite{ref:hortensius1,ref:hortensius2}, Nandi and
Chaudhuri~\cite{ref:nandi,ref:chaudhuri1}, and 
Tomassini et al.~\cite{ref6,ref7,ref8}.
Tomassini has proposed that rule 165 is the best PRNG among the ECA rules.
Moreover, CA with rule 90, rule 105, rule 150, and rule 165 is the best PRNG
among the inhomogeneous CA rules.
This is because Tomassini et al evaluated the randomness using the results of 
the Diehard test suite~\cite{ref:diehard} which does not have a linear
complexity test.
We originally found that these CAs do not pass the linear
complexity test which is one of the NIST statistical tests~\cite{ref:NIST}.
The linear complexity test is crucial for the
application of a PRNG to a cryptosystem because this test detects
whether the prediction is possible. 
It is also known that linear CA is equivalent to the linear feedback
shift register of the same size even if we use these rules 
inhomogeneously~\cite{ref:linear,ref:applied}.
In fact, Nandi and Chaudhuri proposed an additive CA-based block cipher
with nonlinear transformations~\cite{ref:CAC} after realizing this 
point~\cite{ref:comments,ref:reply}.
Mihaljevic and Cattell also independently proposed 
additive CA-based cryptosystems~\cite{ref:M1,ref:M2,ref:Cattell,ref:Cattell2}.

Guan et al. proposed a new class of CA (controllable CA and
two-dimensional CA with an asymmetric neighborship), and investigated their 
randomness using the Diehard test suite~\cite{ref:Guan1,ref:Guan2}.
In this paper, we investigate the randomness of sequences generated by 
nonadditive CAs, that are  ECA30 and its 5-neighbor extension 
(rule 535945230 in 5-neighbor CA framework), using the statistical test suite 
provided by NIST, and compared them with some good PRNGs (AES, SHA1, and MUGI).  
After we show the hardware implementation of these CAs in FPGA, we
demonstrate that these CAs have a high encryption 
speed in experiments of real-time video encryption and decryption systems.

\section{Randomness Evaluation}

Randomness is one of the crucial points for a keystream of secure stream ciphers.
Although various types of statistical test for randomness have been
proposed thus far~\cite{ref:diehard,ref:knuth,ref:fips}, we will focus on the NIST 
statistical test suite~\cite{ref:NIST}, and will show the results of
this test suite.

\subsection{On NIST statistical test suite}

The NIST statistical test suite is a statistical package consisting of 16 tests that
were developed to test the randomness of arbitrary long binary sequences
produced by either hardware or software-based cryptographic random- or
pseudorandom-number generators.
These tests focus on different types of non-randomness that
could exist in a sequence.
The 16 tests are listed in Table 1. 
Note that the test settings of discrete fourier transform test
and Lempel Ziv compression test are wrong~\cite{ref:kim1}.
So, in what follows, we use the corrected version of 
the test suite~\cite{ref:kim2}.

\begin{table}
\begin{center}
\caption{List of the NIST Statistical Tests}
\vspace{1mm}
\begin{tabular}{c|c} \hline \hline
Number & Test Name \\ \hline \hline
1 & Frequency \\ \hline
2 & Block Frequency \\ \hline
3 & Runs \\ \hline
4 & Longest Run \\ \hline
5 & Binary Matrix Rank \\ \hline
6 & Discrete Fourier Transform \\ \hline
7 & Non-overlapping Template Matching  \\ \hline
8 & Overlapping Template Matching  \\ \hline
9 & Universal  \\ \hline
10 & Lempel Ziv Compression \\ \hline
11 & Linear Complexity  \\ \hline
12 & Serial \\ \hline
13 & Approximate Entropy \\ \hline
14 & Cumulative Sums \\ \hline
15 & Random Excursions \\ \hline
16 & Random Excursions Variant \\ \hline
\end{tabular}
\end{center}
\label{table1}
\end{table}

For each statistical test, a set of P-values, which corresponds to
the set of sequences, is produced.
Each sequence is called {\it success} if the corresponding P-value
satisfies the condition P-value $\geq$ $\alpha$,
 and is otherwise called {\it failure}.
For a fixed significance level $\alpha$, $100\alpha$ $\%$ of P-values
are expected to indicate failure\footnote{All the statistical tests of
the NIST statistical test suite have the unique significance level $\alpha=0.01$.}.
For the interpretation of test results, 
NIST adopts the following two approaches, 
\vspace{1mm}

\noindent
(1) the examination of the proportion of success sequences (success rate)
\vspace{1mm}

If the proportion of success sequences falls outside of the following 
    acceptable interval, there is evidence that the data is nonrandom.
\begin{equation}
R \pm 3 \sqrt{\frac{R (1 - R)}{m}} 
\label{eq4}
\end{equation}
Here, $R=1-\alpha$ and $m$ is the number of sequences. 
This interval is determined to be in the 99.73\% range of the normal distribution 
which is an approximation of the binomial distribution under the assumption 
that each sequence is an independent sample.
\vspace{2mm}

\noindent
(2) the examination of the uniformity of the distribution of P-values

This examination is accomplished by computing the following $\chi^{2}$ value.
\begin{equation}
\chi^{2} = \sum_{i=1}^{10} \frac{(F_i - m/10)^{2}}{m/10}
\label{eq5}
\end{equation}
Here, $F_i$ is the number of P-values in subinterval 
[(i-1)*0.1, i*0.1), and $m$ is the number of sequences (sample size).  
The P-value of P-values is calculated such that P$^{\prime}$-value $=$
{\bf igamc} $(9/2, \chi^{2}/2)$, where {\bf igamc}(n,x) is the incomplete gamma
function. 
If P$^{\prime}$-value $\geq$ $0.0001$, then the set of P-values can be
considered to be uniformly distributed.

\subsection{Test results}

In this subsection, we show the results of the NIST statistical 
test suite for several PRNGs.
For each statistical test, the two analyses described above are executed, 
and evaluated whether the set of sequences {\it passes} the test. 
We used 1000 samples of $10^6$ bit sequences for each test. 
Consequently, 10 (keys) $\times$ 1000
(sample) $\times$ $10^6$ (sequence) bits are used for each test in order
to investigate the difference in results between different 
keys\footnote{The key is the initial configuration $\{ S^{t=0}_i \} $ in
the CA case.}.
The input parameters that we used are listed in Table 2.
In the CA case, we used the cell values $\{ S_i^t \}$ with a fixed cell number
$i$ as a keystream, and also used the system size $N=1000$ and 
periodic boundary condition. 
\begin{table}
\begin{center}
\caption{Parameters used for the NIST Test Suite}
\begin{tabular}{c|c} \hline \hline
Test Name & Block Length \\ \hline \hline
Block Frequency & 20,000 \\ \hline
Non-overlapping Template Matching  & 9 \\ \hline
Overlapping Template Matching  & 9 \\ \hline
Universal (Initialization Steps)  & 7 (1280) \\ \hline
Linear Complexity & 500 \\ \hline
Serial & 10 \\ \hline
Approximate Entropy & 10 \\ \hline
\end{tabular}
\end{center}
\label{table2}
\end{table}

\vspace{3mm}
\noindent
{\bf Results of ECA30}
\vspace{2mm}

\begin{table}
\begin{center}
\caption{Results of ECA30.
{\it Pass} denotes a set of sequences that passed all 16 tests.
The other numbers denote the failed test number listed in Table 1.}
\vspace{1mm}
\begin{tabular}{|c|c|c|} \hline \hline
Key & Success Rate  & Uniformity \\ \hline \hline
1 & 3, 7, 15, 16 & pass \\ \hline
2 & 15, 16 & pass \\ \hline
3 & 7  & pass \\ \hline
4 & pass & pass \\ \hline
5 & pass & pass \\ \hline
6 & 7 & pass \\ \hline
7 & pass & pass \\ \hline
8 & 7 & pass \\ \hline
9 & 8 & pass \\ \hline
10 & pass & pass \\ \hline
\end{tabular}
\end{center}
\label{table3}
\end{table}
\begin{figure}
\epsfxsize=8cm 
\epsfysize=8cm 
\centerline{\epsfbox{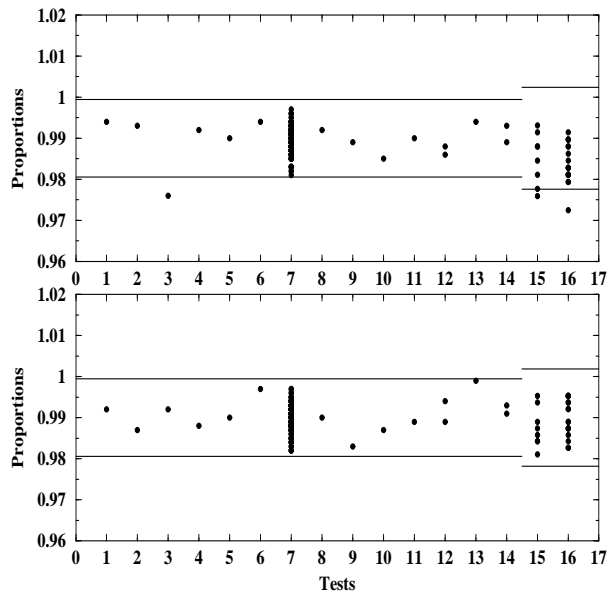}}
\caption{Success rates of ECA30 for 16 tests.
Key 1 and key 4 cases are shown in up and down figures, respectively.
Solid lines denote the acceptable interval (eq.(\ref{eq4}) with $\alpha=0.01$).}
\label{fig:ECA30}
\end{figure}

Table 3 shows the results of ECA 30. 
While all tests are passed in the best cases (key 4, key 5, key 7 and key 10), 
the runs test (number 3), the non-overlapping template matching test
(number 7), the random excursions test (number 15), and the random 
excursion variance test (number 16) fail in the worst case (key 1).
The success rates of the worst case (key 1) and of the best case (key 4)
are shown in Figure \ref{fig:ECA30}. 
Solid lines denote the acceptable interval specified by eq.(\ref{eq4}).
As we can see, some tests have many success rates.
For example, the non-overlapping template matching test (number 7) has
148 success rates because one success rate corresponds to one-template 
(nonperiodic pattern consisting of 9 bits) matching.
If at least one success rates is out of the acceptable interval, then
the test fails (see key 1 case).

\begin{table}
\begin{center}
\caption{Results of ECA30 with rotation shift (11 cells)}
\vspace{1mm}
\begin{tabular}{|c|c|c|} \hline \hline
Key & Success Rate  & Uniformity \\ \hline \hline
1 & pass & pass \\ \hline
2 & pass & pass \\ \hline
3 & 7  & pass \\ \hline
4 & 7 & pass \\ \hline
5 & pass & pass \\ \hline
6 & 7 & pass \\ \hline
7 & pass & pass \\ \hline
8 & 7 & pass \\ \hline
9 & 7 & pass \\ \hline
10 & pass & pass \\ \hline
\end{tabular}
\end{center}
\label{table4}
\end{table}
\begin{figure}
\epsfxsize=8cm 
\epsfysize=8cm 
\centerline{\epsfbox{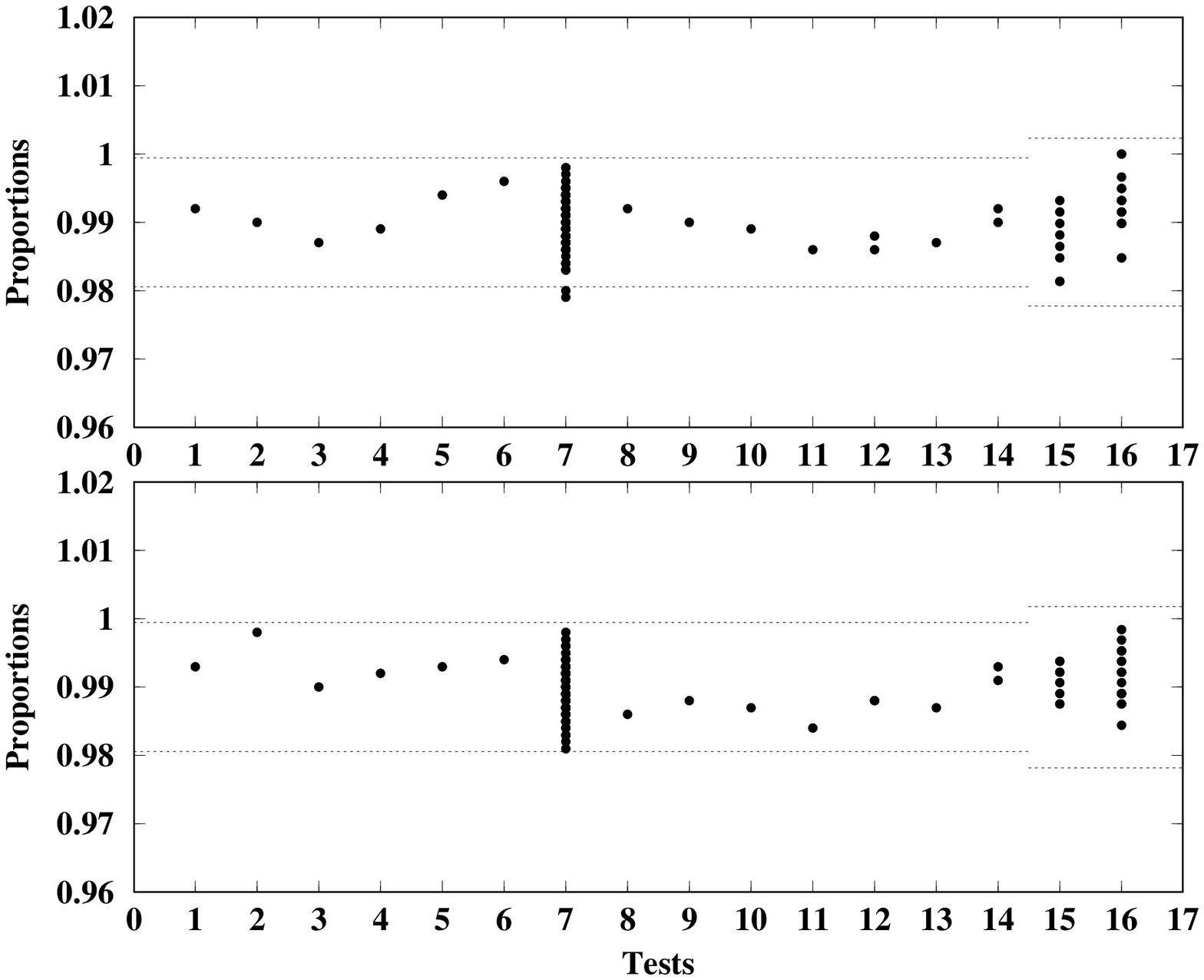}}
\caption{Success rates of ECA30 with rotation shift (11 cells).
Key 9 and key 10 cases are shown in up and down figures, respectively.
Solid lines denote the acceptable interval (eq.(\ref{eq4}) with $\alpha=0.01$).}
\label{fig:ECA30SH11}
\end{figure}

We investigated the test results in the cases that we added rotation shift 
to ECA30 in each time step.
Table 4 shows the results of ECA30 with rotation shift (11 cells). 
The success rates of the worst case (key 9) and of the best case (key 10)
are shown in Figure \ref{fig:ECA30SH11}.
This time, all tests pass in the five cases (key 1, key 2, key 5,
key 7 and key 10).
It seems that the randomness of sequences is slightly improved.  
Although the non-overlapping template matching test fails in
five cases, the number of templates whose success rate is out of 
the acceptable interval (but very close to the boundary) is only one or
two.
However, we found that rotation shift does not always improve 
the randomness of sequences effectively.

\vspace{3mm}
\noindent
{\bf Results of good PRNGs}
\vspace{2mm}

Tables 5, 6, and 7 show the results of AES (128 bit key, OFB mode), 
SHA1, and MUGI, respectively, in order to compare the results 
between ECA30 and them.
\begin{table}
\begin{center}
\caption{Results of AES}
\vspace{1mm}
\begin{tabular}{|c|c|c|} \hline \hline
Key & Success Rate  & Uniformity \\ \hline \hline
1 & pass & pass \\ \hline
2 & pass & pass \\ \hline
3 & 15  & pass \\ \hline
4 & pass & pass \\ \hline
5 & 7 & pass \\ \hline
6 & 14 & pass \\ \hline
7 & 7, 8 & pass \\ \hline
8 & pass & pass \\ \hline
9 & pass & pass \\ \hline
10 & pass & pass \\ \hline
\end{tabular}
\end{center}
\label{table5}
\end{table}
\begin{table}
\begin{center}
\caption{Results of SHA1}
\vspace{1mm}
\begin{tabular}{|c|c|c|} \hline \hline
Key & Success Rate  & Uniformity \\ \hline \hline
1 & pass & pass \\ \hline
2 & pass & pass \\ \hline
3 & 7 & pass \\ \hline
4 & 7 & pass \\ \hline
5 & pass & pass \\ \hline
6 & 7, 15, 16 & pass \\ \hline
7 & 7 & pass \\ \hline
8 & 7 & pass \\ \hline
9 & pass & pass \\ \hline
10 & pass & pass \\ \hline
\end{tabular}
\end{center}
\label{table6}
\end{table}
\begin{table}
\begin{center}
\caption{Results of MUGI}
\vspace{1mm}
\begin{tabular}{|c|c|c|} \hline \hline
Key & Success Rate  & Uniformity \\ \hline \hline
1 & 7 & pass \\ \hline
2 & pass & pass \\ \hline
3 & pass & pass \\ \hline
4 & pass & pass \\ \hline
5 & 7 & pass \\ \hline
6 & pass & pass \\ \hline
7 & pass & pass \\ \hline
8 & pass & pass \\ \hline
9 & 7 & pass \\ \hline
10 & pass & pass \\ \hline
\end{tabular}
\end{center}
\label{table7}
\end{table}
As we can see, all tests are passed in six cases (AES), in five cases
(SHA1), and in seven cases (MUGI), respectively.
Note that the SHA1 case is the same frequency as the ECA30 
with rotation shift (11 cell).

\vspace{3mm}
\noindent
{\bf Results of 5-neighbor CA}
\vspace{2mm}

We can obtain the following equation if we consider two iterations of
eq.(\ref{eq2}),
\begin{eqnarray}
S_i^{t+1} & = & S_{i-2}^t \oplus S_{i+1}^t \oplus S_{i+2}^t \oplus \nonumber \\ 
 & & S_{i-1}^t \cdot S_{i+1}^t  \oplus S_{i-1}^t \cdot S_{i+2}^t \oplus S_{i}^t \cdot S_{i+1}^t    \oplus \nonumber \\
 & & S_{i}^t \cdot S_{i+2}^t \oplus S_{i+1}^t \cdot S_{i+2}^t \oplus \\
 & & S_{i-1}^t \cdot S_{i+1}^t \cdot S_{i+2}^t      
\oplus S_{i}^t \cdot S_{i+1}^t \cdot S_{i+2}^t \nonumber
\end{eqnarray}
This is equivalent to rule 535945230 in the 5-neighbor CA framework~\cite{ref:new}.
We have investigated the randomness of sequences generated by some class
of 5-neighbor CA rules.
We found that rule 535945230 is the best.

Table 8 show the results of CA5-535945230 with rotation shift (11 cells).
We use one cell {$S_i^t$} (fixed $i$) as a keystream at each time step as well as
ECA30 cases.
As we can see, all tests are passed in six cases.
This is the same frequency as AES.
We can conclude that the CA5-535945230 with rotation shift (11 cells) 
has good randomness, which is comparable to well-known good PRNGs 
such as AES, SHA1, and MUGI.  

\begin{table}
\begin{center}
\caption{Results of CA5-535945230 with rotation shift (11 cells)}
\vspace{1mm}
\begin{tabular}{|c|c|c|} \hline \hline
Key & Success Rate  & Uniformity \\ \hline \hline
1 & pass & pass \\ \hline
2 & 7, 12 & pass \\ \hline
3 & pass & pass \\ \hline
4 & pass & pass \\ \hline
5 & 7 & pass \\ \hline
6 & pass & pass \\ \hline
7 & pass & pass \\ \hline
8 & 7 & pass \\ \hline
9 & pass & pass \\ \hline
10 & 7 & pass \\ \hline
\end{tabular}
\end{center}
\label{table8}
\end{table}

\subsection{Security discussion}

It is known that ECA30-based stream cipher which was proposed by Wolfram
has a security problem.
If we use two consecutive cell values ($S_i, S_{i+1}$) as a keystream 
at each time step, an attacker can easily calculate the secret
key (initial configuration) from the keystream using the following 
equation which is the same equation as eq.(\ref{eq2}). 
\begin{equation}
S_{i-1}^t = S_{i}^{t+1} \oplus S_i^t \oplus S_{i+1}^t \oplus S_i^t 
\cdot S_{i+1}^t
\label{eqkai}
\end{equation}
In order to avoid this, Wolfram proposed that we should use only one cell
value ($S_i$) as a keystream at each time step.
He suggested that an attacker cannot easily calculate the secret key
from the keystream in this case (exponential time is required).  
However, the effective key size is much less than $N$ even
in this case~\cite{ref:analysis}.
We should set system size $N=2000$ in order to set the effective key
size to more
than $80$ in this Wolfram case.

\begin{figure}
\epsfxsize=6cm 
\epsfysize=5cm 
\centerline{\epsfbox{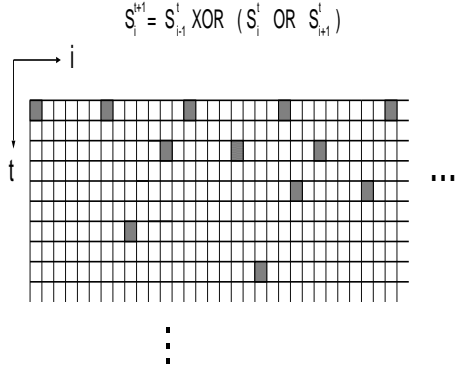}}
\caption{Sampled cells in CA5-535945230 case}
\label{fig:shasenCA5}
\end{figure}

In order to avoid this attack, we sample cell values such that
the distance between consecutively sampled cells becomes larger 
(e.g., cell numbers 1, 7, 14, 22, 31, 41, $\cdots$, 932, 976 are sampled for
40 bit per clock), and rotation shift (11 cells) is added 
at each time step.
As a result, we sample cell values which are denoted as shaded cells 
in Figure \ref{fig:shasenCA5} in the CA5-535945230 case if we consider 
the 3-neighbor CA framework.
In this case, the attack mentioned above no longer applies directly. 
It is difficult to calculate the secret key from the keystream (shaded cells) 
using eq.(\ref{eqkai}).
If someone could find another attack, the effective key size would be
improved as compared with the Wolfram type.

\begin{figure}
\epsfxsize=8cm 
\epsfysize=8cm 
\centerline{\epsfbox{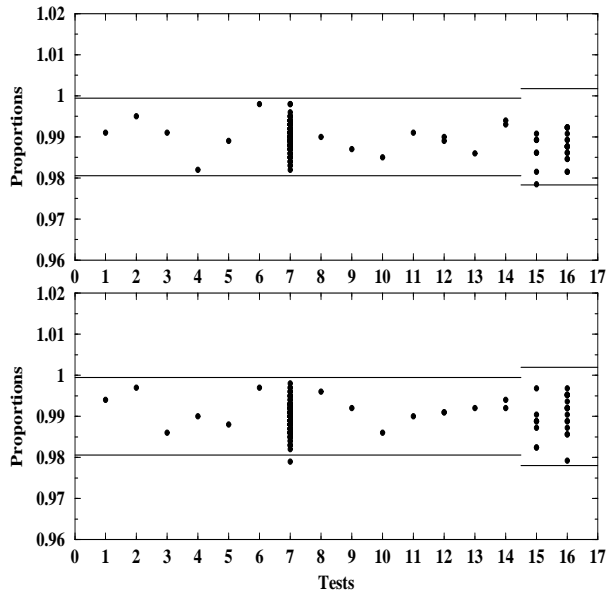}}
\caption{Success rates of ECA30 with sampling method for two different
 keys}
\label{fig6}
\end{figure}

\begin{figure}
\epsfxsize=8cm 
\epsfysize=8cm 
\centerline{\epsfbox{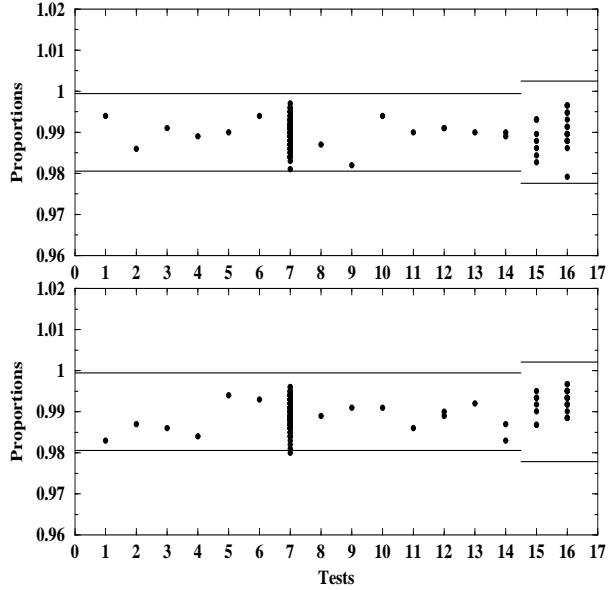}}
\caption{Success rates of CA5-535945230 with sampling method for two different keys}
\label{fig7}
\end{figure}

The keystream using this sampling method also has high-quality
randomness.
The success rates of the ECA30 case and CA5-535945230 case are shown 
in Figure \ref{fig6} and Figure \ref{fig7}, respectively.
As we can see, all tests are passed except the non-overlapping template
matching test (for one template).
Note that the linear complexity test is also passed even if we choose 
the maximum parameter $M=5000$ in both cases.

It is well known that statistical characteristics of the keystream are
just a component of the security evaluation of a stream cipher.
In this paper, we propose an encryption approach based on CA with desirable
statistical characteristics and implementation suitability, but that its
detailed security evaluation is out of our scope (and that this
issue is an open one).

\section{Hardware Implementation}

\begin{figure}
\epsfxsize=6cm 
\epsfysize=6cm 
\centerline{\epsfbox{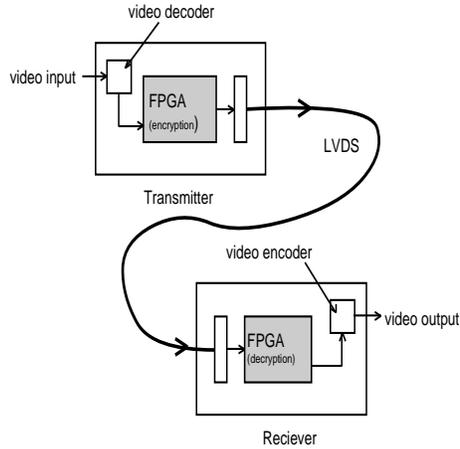}}
\caption{DDR-SDRAM evaluation boards}
\label{board}
\end{figure}

\begin{figure}
\epsfxsize=10cm 
\epsfysize=6cm 
\centerline{\epsfbox{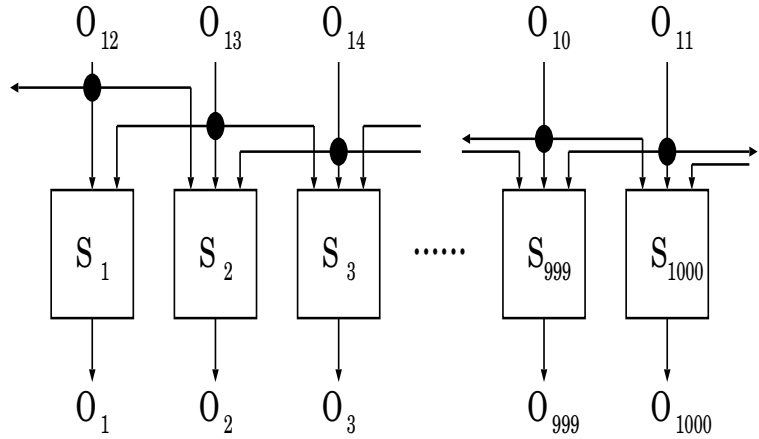}}
\caption{CA-based PRNG core}
\label{block}
\end{figure}

\begin{figure}
\epsfxsize=8cm 
\epsfysize=8cm 
\centerline{\epsfbox{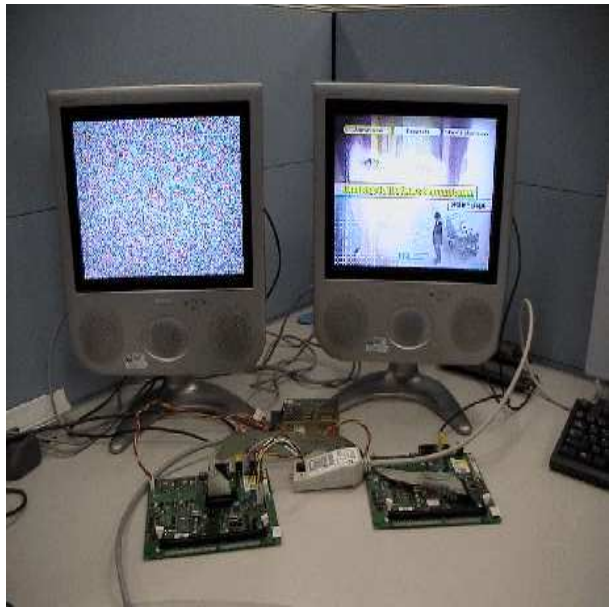}}
\caption{Real-time video encryption and decryption system}
\label{experi}
\end{figure}

Figure \ref{board} shows a schematic of DDR-SDRAM evaluation boards produced by Tokyo
Electron Device Ltd.
There are video input and output (40 bit per clock), FPGA (VirtexII),
and low voltage differential signaling (LVDS) on this board 
for the purpose of real-time video encryption and decryption. 
We implemented the CA-based stream cipher on two FPGAs 
(see Fig. \ref{block}), and executed the
experiment of real-time video encryption and decryption (see Fig. \ref{experi}).

\subsection{Implementation results}

Table 9 shows the implementation results for the system size $N=1000$ case.
\begin{table}
\begin{center}
\caption{Implementation results}
\vspace{1mm}
\begin{tabular}{c|c|c} \hline \hline
 & ECA30 & CA5 \\ \hline 
gate size (gate) & 14699 &  20699       \\ \hline 
max clock frequency (MHz) & 105.83 & 75.55    \\ \hline
encryption speed (Gbps) & 4.23 & 3.02   \\ \hline
\end{tabular}
\end{center}
\end{table}
As we can see, both algorithms work up to a high clock frequency 
because of their simple construction.
In the CA5 case, randomness and security level are higher than those in
the ECA30 case although encryption speed and gate size are lower.  
If we set the system size $N$ larger, the encryption speed becomes higher.
On the other hand, if we set the system size $N$ smaller, the gate size 
becomes smaller although we have shown only the system size $N=1000$ case.
Actually, we realized a 1 Gbps encryption speed because the board we used
in the real-time encryption and decryption experiment 
has a 27 MHz clock frequency.

\section{Summary}

We have reviewed the cellular automaton-based PRNGs, and 
have shown that one of these PRNGs, which was denoted as
CA5-535945230, can generate high-quality random numbers 
which can pass all of the NIST statistical tests. 
We demonstrated that the encryption algorithm using the CA-based PRNG 
has a 3 Gbps encryption speed in the case of FPGA (20 Kgate used).
This suggests that CA is suitable for developing a high-speed
hardwarelike encryption system.

\section*{Acknowledgements}

The corresponding author acknowledges a research fellowship from the 
Japan Society for the Promotion of Science for Young Scientists (JSPS).



\vspace{5mm}
\noindent (Received May 14, 2004; revised September 7, 2004)


\begin{thebibliography}{99}
%
\bibitem{ref:handbook}
Menezes et al.: Handbook of Applied Cryptography, CRC Press, 1997.
%
\bibitem{ref:CHES}
T.~E.~Tkacik: A hardware random number generator, Lecture Notes in Computer
        Science, Vol. 2523 (CHES2002), pp. 450--453.
%
\bibitem{ref:wolfram1}
S.~Wolfram: Random sequence generation by cellular automata,
Advances~in~Applied~Mathematics, Vol. 7, pp. 123--169, 1986.
%
\bibitem{ref:wolfram2}
S.~Wolfram: Cryptography with cellular automata, 
Lecture Notes in Computer Science, Vol. 0218 (CRYPTO'85), pp. 429--432.
%
\bibitem{ref:new}
S.~Wolfram: A New Kind of Science, Wolfram Media, Inc., 2002.
%
\bibitem{ref:hortensius1}
P.~D.~Hortensius et al.: Parallel random number generation for VLSI
        systems using cellular automata, IEEE Transactions on Computers, 
	Vol. 38, pp. 1466--1473, 1986.
%
\bibitem{ref:hortensius2}
P.~D.~Hortensius et al.: Cellular automata-based pseudorandom number
        generators for built-in self-test, IEEE Transactions on
        Computer-aided Design, Vol. 8, pp. 842--859, 1989.
%
\bibitem{ref:nandi}
S.~Nandi et al.: Theory and applications of cellular automata in
        cryptography, IEEE Transactions on Computers, Vol. 43,
	pp. 1346--1357, 1994.
%
\bibitem{ref:chaudhuri1}
P.~Chaudhuri et al.: Additive Cellular Automata, IEEE Computer
        Society Press, 1997.
%
\bibitem{ref6}
M.~Sipper and M.~Tomassini: Generating parallel random number
        generators by cellular programming, Int.~J.~Mod.~Phys., Vol. 7,
	pp. 181--190, 1996.
%
\bibitem{ref7} 
M.~Tomassini et al.: Generating high-quality random numbers in
        parallel by cellular automata, Future Generation Computer
	Systems, Vol. 16, pp. 291--305, 1999.
%
\bibitem{ref8}
M.~Tomassini and M.~Perrenoud: Nonuniform cellular automata for
        cryptography, Complex Systems, Vol. 12, pp. 71--81, 2000.
%
\bibitem{ref:diehard}
G.~Marsagglia: Diehard Test,\\
http://stat.fsu.edu/$\sim$geo/diehard.html, 1998.
%
\bibitem{ref:NIST}
A.~Rukhin, et al., A Statistical Test Suite for Random and Pseudorandom Number Generators 
for Cryptographic Applications, NIST, http://csrc.nist.gov/rng/, 2001.
%
\bibitem{ref:linear}
P.~H.~Bardell: Analysis of cellular automata used as pseudorandom
        pattern generators, International Test Conference,
	pp. 762--768, 1990.
%
\bibitem{ref:applied}
B.~Schneier: Applied Cryptography (second ed.), John Wiley \& Sons,
        Inc., pp. 414, 1996.
%
\bibitem{ref:CAC}
S.~Sen et al.: Cellular automata based cryptosystem (CAC),
        Lecture Notes in Computer Science, Vol. 2513 (ICICS2002), pp. 303--314.
%
\bibitem{ref:comments}
S.~R~.Blackburn et al.: Comments on theory and applications of
        cellular automata in cryptography, IEEE Transactions on
        Computers, Vol. 46, pp. 637--638, 1997.
%
\bibitem{ref:reply}
S.~Nandi and P.~Chaudhuri: Reply to comments on theory and applications of
        cellular automata in cryptography, IEEE Transactions on
        Computers, Vol. 46, pp. 638--639, 1997.
%
\bibitem{ref:M1}
M.~J.~Mihaljevic: An improved key stream generator based on the
	programmable cellular automata, Lecture Notes in Computer
	Science, Vol. 1334, pp. 181--191, 1997.
%
\bibitem{ref:M2}
M.~J.~Mihaljevic and H.~Imai: A family of fast keystream generators
	based on programmable linear cellular automata over GF(q) and
	time-variant table, IEICE Transactions on Fundamentals,
	Vol. E82-A, pp. 32--39, 1999.
%
\bibitem{ref:Cattell}
K.~Cattell and J.~C.~Muzio: Synthesis of one-dimensional linear hybrid
	cellular automata, IEEE Transactions on Computer-Aided Design of
	Integrated Circuits and Systems, Vol. 15, pp. 325--335, 1996.
%
\bibitem{ref:Cattell2}
K.~Cattell et al.: 2-by-n hybrid cellular automata with regular
	configuration: theory and application, IEEE Transactions on
	Computers, Vol. 48, pp. 285--295, 1999.
%
\bibitem{ref:Guan1}
S.-U. Guan and S. Zhang: An evolutionary approach to the design
      of controllable cellular automata structure for random number
      generation, IEEE Trans. on Evolutionary Computation, Vol. 7 
      No. 1, pp. 23--36, 2003.
%
\bibitem{ref:Guan2}
S.-U. Guan, S. Zhang and M.T. Quieta: 2-D CA variation with
      asymmetric neighborship for pseudorandom number generation,
      IEEE Trans. Comput. Aid. Design, Vol. 23, pp. 378--388, 2004.
%
\bibitem{ref:knuth}
D.~Knuth: Seminumerical Algorithms, Addson-Wesley, Reading, Mass., 1981.
%
\bibitem{ref:fips}
Security Requirements for Cryptographic Modules, NIST,\\ 
http://csrc.nist.gov/publications/fips/fips140-2/fips1402.pdf, 2001.
%
\bibitem{ref:kim1}
S.~J.~Kim, K.~Umeno and A.~Hasegawa : Validity Check of NIST 800-22 
	Statistical Test Suite for Randomness, 
ISM Report on Research and Education, No.17, pp. 328--329, 2003.
%
\bibitem{ref:kim2}
S.~J.~Kim, K.~Umeno and A.~Hasegawa : On the NIST Statistical Test
        Suite for Randomness, Technical Report of IEICE Vol.103, No.499 
(ISEC2003-87), pp.21--27,\\
http://www.iacr.org/2004/018.pdf.
%
\bibitem{ref:analysis}
W.~Meier and O.~Staffelbach: Analysis of pseudo random sequences
        generated by cellular automata, Lecture Notes in Computer
        Science, Vol. 0547, pp. 186--199, 1991.
%
\end{thebibliography}
\end{document}